\documentclass[archive,final,onefignum,onetabnum]{siamart220329}

\usepackage{braket,amsfonts}

\usepackage{array}

\usepackage[caption=false]{subfig}
\usepackage{pgfplots}

\newsiamthm{claim}{Claim}
\newsiamremark{remark}{Remark}
\newsiamremark{hypothesis}{Hypothesis}
\crefname{hypothesis}{Hypothesis}{Hypotheses}

\usepackage{algorithmic}

\usepackage{graphicx,epstopdf}

\Crefname{ALC@unique}{Line}{Lines}

\usepackage{amsopn}

\usepackage{amsmath,amssymb,amsfonts}%
\usetikzlibrary{arrows.meta}
\newcommand{\bsym}{\boldsymbol} 
\newcommand{\mbf}{\mathbf}  
\newcommand{\tss}{\textsuperscript}  
\def\Tr{^\mathsf{T}} 

\title{One-shot Parareal Approach for Topology Optimisation of Transient Heat Flow \thanks{Submitted to the editors September 24, 2024.
\funding{The work has been funded by Independent Research Foundation Denmark (DFF) through a Sapere Aude Research Leader grant (3123-00020B) for the COMFORT project (COmputational Morphogenesis FOR Time-dependent problems).}}}

\author{Magnus Appel\thanks{Institute of Mechanical and Electrical Engineering, University of Southern Denmark, Odense, 5230, Denmark (\email{magap@sdu.dk)} } 
\and Joe Alexandersen\thanks{Institute of Mechanical and Electrical Engineering, University of Southern Denmark, Odense, 5230, Denmark (\email{joal@sdu.dk)}} }

\headers{One-shot Parareal Method of Topology Optimisation}{Magnus Appel and Joe Alexandersen}

\ifpdf
\hypersetup{ pdftitle={Topology Optimisation with Parareal} }
\fi

\begin{document}
\maketitle

\begin{abstract}
This paper presents a method of performing topology optimisation of transient heat conduction problems using the parallel-in-time method Parareal. To accommodate the adjoint analysis, the Parareal method was modified to store intermediate time steps. Preliminary tests revealed that Parareal requires many iterations to achieve accurate results and, thus, achieves no appreciable speedup. To mitigate this, a one-shot approach was used, where the time history is iteratively refined over the optimisation process. The method estimates objectives and sensitivities by introducing cumulative objectives and sensitivities and solving for these using a single iteration of Parareal, after which it updates the design using the Method of Moving Asymptotes. The resulting method was applied to a test problem where a power mean of the temperature was minimised. It achieved a peak speedup relative to a sequential reference method of $5\times$ using 16 threads. The resulting designs were similar to the one found by the reference method, both in terms of objective values and qualitative appearance. The one-shot Parareal method was compared to the Parallel Local-in-Time method of topology optimisation. This revealed that the Parallel Local-in-Time method was unstable for the considered test problem, but it achieved a peak speedup of $12\times$ using 32 threads. It was determined that the dominant bottleneck in the one-shot Parareal method was the time spent on computing coarse propagators. 
\end{abstract}




\section{Introduction}

Topology optimisation is a branch of mathematical design optimisation which aims to find the optimal material distribution of a given structure / device for a given purpose. It has been applied to a variety of problems \cite{Deaton2013_survey}, including problems involving heat transfer \cite{Dbouk2017_review} which will be the focus of this paper. In the context of transient heat transfer, it has been demonstrated that transient effects can have a noticeable effect on the optimised designs \cite{wu2019_minmaxTemp,Zeng2020_transForcedConv}. 

In general, performing topology optimisation is very computationally expensive since the state of the system (plus the corresponding adjoint state in the case of adjoint methods) has to be solved for at each iteration of the optimisation method. Moreover, topology optimisation of transient problems is even more expensive than their stationary counterparts, since the computational cost is multiplied by the number of time steps.


Several approaches have been employed in order to accelerate the computations involved in topology optimisation methods. One such approach is to use iterative solvers to solve for the states and adjoint states of the system, while only performing one or a few iterations of the iterative solvers during each optimisation cycle. Such an approach is called a ``one-shot" approach and was initially proposed as a way of solving optimal control problems \cite{oneShotOrig}. Since then, it has also been used to solve topology optimisation problems \cite{Laniewski2016_oneShot2016,amir2024_OneShot}. 


Another way to accelerate topology optimisation methods is by parallelising the computation with respect to the spatial domain using domain decomposition. This approach has been used to solve various large scale topology optimisation problems \cite{Borrvall2001_largeScaleTopOpt,aage2015_openSourceTopOpt,Alexandersen2016_largeScaleTopOpt,aage2017_gigavoxel}. When it comes to solving transient problems, it can be noted that this method of parallelising the process with respect to space has a natural limitation, in the sense that even if 100\% parallel efficiency is achieved, the computational cost will still be proportional to the number of time steps. As such, it would be convenient to parallelise the process with respect to the time axis. 


Despite the fact that time-dependent systems are intuitively thought of as being sequential in nature, parallel-in-time methods of solving initial value problems (IVPs) do exist. One such method is the ``Parareal" method \cite{parareal_orig}, which is an iterative, non-intrusive parallel-in-time method. It is non-intrusive in the sense that it works by making calls to existing time-stepping methods. Specifically, it makes calls to an expensive time-stepping method in parallel, after which it applies a sequential correction to the solution using a cheaper time-stepping method. 

Another parallel-in-time method is MultiGrid Reduction In Time (MGRIT) \cite{mgrit_original}. This is a multigrid method where the time-axis is treated as being the grid, and the coarse grids are constructed by coarsening the time-axis. It has been proven that Parareal is equivalent to two-level MGRIT in the case of specific choices of relaxation methods, restriction operators, and prolongation operators used by MGRIT \cite{Gander2007_pararealAnalysis}. Other parallel-in-time methods have also been devised, such as space-time multigrid methods \cite{horton1995_stMultigrid,Steinbach2018_FE_ST_MG} and PFASST \cite{pfasst_original}. 


Various parallel-in-time methods have been applied to solve optimal control problems, including methods involving Parareal-based preconditioners \cite{du2013_pararealPrecond,Ulbrich2015_pararealProcond} and PFASST \cite{Minion2018_pfasstOptCtrl}. In addition, a non-intrusive one-shot approach employing MGRIT has been applied to solve optimal control problems \cite{Gunther2019_mgritOptCtrl}. A method named ``ParaOpt" has been proposed, which is a Newton-based parallel-in-time method of solving optimisation problems \cite{Gander_paraopt}. Recently, a parallel-in-time method named ``Parallel Local-in-Time" (PLT) has been proposed and applied specifically for the purpose of performing topology optimisation \cite{Theulings_PLT}. 
\\
\\
This paper proposes a one-shot method using Parareal to accelerate topology optimisation of transient heat conduction problems. The paper is structured as follows: \Cref{sec: gov eqs} defines the governing equations of interest; \Cref{sec: prob statement} defines the problem statement; \Cref{sec: num methods} explains the sensitivity analysis and modifications to Parareal to accommodate the sensitivity analysis; \Cref{sec: prelim tests} presents preliminary experiments regarding the speed and accuracy of the proposed method, which form the motivation for employing a one-shot method; \Cref{sec: one-shot approach} presents the one-shot Parareal method and experimental results of it; \Cref{sec: PLT comparison} presents an experimental comparison between the Parallel Local-in-Time method and the one-shot Parareal method; and \Cref{sec: conclusion} gives a conclusion.


\section{Governing equations}
\label{sec: gov eqs}


This paper considers transient heat conduction in two dimensions governed by the following partial differential equation and boundary conditions:
\begin{subequations}
\begin{align}
    c \frac{\partial T}{\partial t} - \bsym{\nabla} \cdot (k \bsym{\nabla} T) &= q  \quad \text{on } \Omega \text{ for } 0 \leq t \leq t_T
    \\
    T &= 0 \quad  \text{on } \Omega \text{ at } t=0
    \\
    T &= 0 \quad  \text{on } \Gamma_D
    \\
    \hat{\mbf{n}} \cdot \bsym{\nabla} T &= 0  \quad \text{on }  \Gamma_N
\end{align}
\end{subequations}
where $T=T(\mbf{x},t)$ is the temperature field, $c=c(\mbf{x})$ is the volumetric heat capacity, $k=k(\mbf{x})$ is the thermal conductivity, $q=q(\mbf{x},t)$ is the imposed external heat load per unit volume, $\Omega$ is the two-dimensional spatial domain, $t_T$ is the terminal time of the problem, $\Gamma_D$ is the subset of the boundary where Dirichlet boundary conditions are imposed, $\Gamma_N$ is the subset of the boundary where Neumann boundary conditions are imposed, and $\hat{\mbf{n}}$ is the unit normal vector on the boundary. 

\subsection{Design parametrisation}

The spatial domain is divided into two subdomains: $\Omega_C$, which is filled with one material with a high conductivity; and $\Omega \setminus \Omega_C$ which is filled with a different material with a low conductivity. The design field associated with these subdomains is defined as:
\begin{align}
    \chi(\mathbf{x}) &= 
    \begin{cases}
    1 & \text{for } \mathbf{x} \in \Omega_C \\ 
    0 & \text{for } \mathbf{x} \notin \Omega_C 
    \end{cases}
\end{align}
To enable the use of gradient-based methods for topology optimisation, the requirement that $\chi(\mathbf{x})$ must be either 0 or 1 will be relaxed and instead the condition that $0 \leq \chi(\mathbf{x}) \leq 1$ is imposed. 

The problem is discretised in space using element-wise constant functions for $k$, $c$, and $\chi$. To encourage black-and-white solutions ($\chi\in\left\lbrace 0,1 \right\rbrace$), a threshold projection filter \cite{guest2004_projOrig} is applied to the discretised version of $\chi$ by first applying a linear filter operator, $H$, to the design field:
\begin{align}
    \chi_{fil} &= H \{ \chi \} 
\end{align}
The linear filter used in this work computes a weighted average of the design field within a distance of $r_{fil}$ around each point, where $r_{fil}$ is called the filter radius \cite{bourdin2001_topOptFilter,bruns2001_topOptFilter}. After applying the linear filter, the following non-linear projection function \cite{wang2011_projAndRobust} is applied to $\chi_{fil}$:
\begin{align}
    \chi_{phys} &= P( \chi_{fil} )
    =
    \frac{ \tanh\left[ \beta \cdot ( \chi_{fil} - \eta ) \right] + \tanh\left[ \beta \eta \right]  }{ \tanh\left[ \beta \cdot ( 1 - \eta ) \right] + \tanh\left[ \beta \eta \right] } 
\end{align}
where $\eta$ is a threshold parameter and $\beta$ is a parameter which controls the ``sharpness" of the output. This ``physical design field", $\chi_{phys}$, is then what determines the material parameters $c$ and $k$. 

To interpolate between the properties of the conducting and insulating material, the following Solid Isotropic Material with Penalisation (SIMP) scheme is used to define $c$ and $k$ at every point:
\begin{align}
    c(\mbf{x}) = 
    c\left( \chi_{phys}(\mbf{x}) \right) 
    &= 
    c_{min} + (c_0 - c_{min}) {\chi_{phys}}^{p_c} 
    \\
    k(\mbf{x}) = 
    k\left(\chi_{phys}(\mbf{x}) \right) 
    &= 
    k_{min} + (k_0 - k_{min}) {\chi_{phys}}^{p_k} 
\end{align}
where $c_0$ and $k_0$ are the material properties of the conductor, $c_{min}$ and $k_{min}$ are the material properties of the insulator, and $p_c$ and $p_k$ are penalty powers assigned to the heat capacity and conductivity, respectively. 

\subsection{Discretisation}

The problem is discretised in space using a simple Galerkin finite element method with rectangular elements with bilinear shape functions for $T$ and element-wise constant functions for $q$. The first-order backward Euler method is used as the time-stepping method and the solution is evaluated at time points $t_0, t_1, \hdots t_{N_t}$, where $N_t$ is the desired number of time steps. The time points are set to be evenly spaced, so $t_n \equiv n \cdot \Delta t$ where $\Delta t \equiv t_T / N_t$. As such, the governing equations of the discretised problem are of the following form:
\begin{subequations}
\begin{align}
    \mbf{T}_0 &= \mbf{0}
    \\
    \label{eq: euler method for T}
    \mbf{C} \frac{\mbf{T}_n - \mbf{T}_{n-1}}{\Delta t} + \mbf{K} \mbf{T}_n &= \mbf{q}_n
    \quad \text{for } n=1,2, \hdots N_t-1, N_t
\end{align}
\end{subequations}
where $\mbf{T}_n$ is the vector containing the nodal values of the temperature at the time $t_n$, $\mbf{C}$ is the heat capacity matrix, $\mbf{K}$ is the thermal conductivity matrix, and $\mbf{q}_n$ is the external heat load vector at $t_n$. The above discretised IVP is referred to as the primal problem. These governing equations can be written in all-at-once matrix form like so:
\begin{align}
    \label{eq: all-at-once matrix form long}
    \begin{pmatrix}
        \frac{\mbf{C}}{\Delta t} + \mbf{K} & \mbf{0} & \mbf{0} & \hdots & \mbf{0} & \mbf{0} \\
        -\frac{\mbf{C}}{\Delta t}  & \frac{\mbf{C}}{\Delta t} + \mbf{K} & \mbf{0} & \hdots & \mbf{0} & \mbf{0} \\
        \mbf{0} & -\frac{\mbf{C}}{\Delta t}  & \frac{\mbf{C}}{\Delta t} + \mbf{K} & \hdots & \mbf{0} & \mbf{0} \\
        \vdots & \vdots & \vdots & \ddots & \vdots & \vdots \\
        \mbf{0} & \mbf{0} & \mbf{0} & \hdots & \frac{\mbf{C}}{\Delta t} + \mbf{K} & \mbf{0} \\
        \mbf{0} & \mbf{0} & \mbf{0} & \hdots & -\frac{\mbf{C}}{\Delta t} & \frac{\mbf{C}}{\Delta t} + \mbf{K}
    \end{pmatrix}
    \begin{pmatrix}
        \mbf{T}_1 \\ \mbf{T}_2 \\ \mbf{T}_3 \\ \vdots \\ \mbf{T}_{N_t-1} \\ \mbf{T}_{N_t}
    \end{pmatrix}
    &=
    \begin{pmatrix}
        \mbf{q}_1 \\ \mbf{q}_2 \\ \mbf{q}_3 \\ \vdots \\ \mbf{q}_{N_t-1} \\ \mbf{q}_{N_t}
    \end{pmatrix}
\end{align}
This is compacted down to the following form:
\begin{align}
    \label{eq: all-at-once matrix form short}
    \mbf{M} \, \mbf{S} &= \mbf{Q}
\end{align}
where $\mbf{S} = \left( \mbf{T}_1\Tr \,\, \mbf{T}_2\Tr \,\, \hdots \,\, \mbf{T}_{N_t}\Tr \right) \Tr$, $\mbf{Q} = \left( \mbf{q}_1\Tr \,\, \mbf{q}_2\Tr \,\, \hdots \,\, \mbf{q}_{N_t}\Tr \right) \Tr$, and  $\mbf{M}$ is the square matrix which appears in \Cref{eq: all-at-once matrix form long}.

For this study, the objective functional considered is intended to be very sensitive to the details of the time-evolution of the temperature. To achieve this, the objective functional is set to be a power mean of the temperature over both space and time:
\begin{align}
    \Theta &= \left( \frac{1}{t_T A} \int_{0}^{t_T} \iint_\Omega T^p \, \mathrm{d}^2 \mbf{x} \, \mathrm{d} t \right)^{1/p}
\end{align}
where $A$ is the area of $\Omega$ and $p$ is a constant which is set to 20 in this work. As such, this objective function can be thought of as an approximation of the max-norm (i.e. infinity norm) of the temperature. In the case of evenly spaced elements and evenly spaced time points, it is reasonable to approximate this objective functional as the following in the context of the discretised temperature:
\begin{align} \label{eq:objective_discrete}
    \Theta &= \left( \frac{1}{N_t N_e} \sum_{j=1}^{N_t} {\lVert \mbf{T}_j \rVert_p}^p  \right)^{1/p}
\end{align}
where $N_e$ is the number of elements and $\lVert \mbf{T}_j \rVert_p$ is the $p$-norm of $\mbf{T}_j$ (note that the above equation only applies if $p$ is even, which it is in this case). The above formula can also be expressed more compactly in terms of $\mbf{S}$ as the following:
\begin{align} \label{eq:objective_simple}
    \Theta &= \left( N_t N_e \right)^{-1/p} \lVert \mbf{S} \rVert_p
\end{align}


\section{Problem statement}
\label{sec: prob statement}


Using the notation established in \Cref{sec: gov eqs}, the topology optimisation problem of interest in this paper may be formulated as:
\begin{subequations} \label{eq:optprob}
\begin{align}
    \min_{\boldsymbol{\chi}, \mbf{S}} \quad & \Theta = \left( N_t N_e \right)^{-1/p} \lVert \mbf{S} \rVert_p
    \\
    \textrm{s.t.} \quad 
    & \mbf{M} \, \mbf{S} = \mbf{Q} \label{eq:optprob-physics}
    \\
    & 0 \leq \chi(\mathbf{x}) \leq 1 \quad \forall  \, \mathbf{x} \in \Omega 
    \\
    & \iint_\Omega \chi_{phys}(\mathbf{x}) \, \mathrm{d}^2\mathbf{x} \leq a_{max} A 
\end{align}
\end{subequations}
where $\boldsymbol{\chi}$ is the vector containing the degrees of freedom of $\chi(\mbf{x})$ and $a_{max}$ is the largest desired area fraction to be occupied by the conducting material. In addition to solving the above topology optimisation problem, this work aims to use a parallel-in-time method named ``Parareal" to speed up the optimisation process.


\section{Numerical methods}
\label{sec: num methods}



\subsection{Adjoint method of sensitivity analysis}
\label{sec: adj method}


The topology optimisation problem in \Cref{eq:optprob} is solved using either: a common nested approach, where the physics in \Cref{eq:optprob-physics} is solved ``exactly'' for every step of the optimisation algorithm; a one-shot approach, where the physics is iteratively refined over the optimisation process. As the optimisation algorithm, a version of the Method of Moving Asymptotes (MMA) \cite{MMA_original} is used,  modified to work well even when the sharpness parameter, $\beta$, is large in the beginning of the optimisation process \cite{guestConstBeta}. This is a gradient-based method, meaning that the gradient, or sensitivities, of the objective, $\Theta$, with respect to the design field, $\bsym{\chi}$, must be computed. A formula for the sensitivities can be found by first defining the residual to be $\mbf{R} \equiv \mbf{M} \, \mbf{S} - \mbf{Q}$, after which the sensitivities can be expressed as being the total derivative of $\Theta$ with respect to $\bsym{\chi}$ under the constraint that $\mbf{R} = \mbf{0}$. This total derivative is given by the following formula:
\begin{align}
    \bsym{\nabla} \Theta
    = 
    \frac{\mathrm{d} \Theta}{\mathrm{d} \bsym{\chi}} 
    &= 
    \frac{\partial \Theta}{\partial \bsym{\chi}} 
    - \frac{\partial \Theta}{\partial \mbf{S}} \left( \frac{ \partial\mbf{R} }{ \partial\mbf{S} } \right)^{-1} \frac{\partial \mbf{R}}{ \partial \bsym{\chi} }
\end{align}
For the considered problem, $ \bsym{\nabla} \Theta$ reduces to:
\begin{align} \label{eq:sensitivity}
    \bsym{\nabla} \Theta
    &=
    - \frac{\partial \Theta}{\partial \mbf{S}} \mbf{M}^{-1} \frac{\partial \mbf{R}}{ \partial \bsym{\chi} }
\end{align}
The sensitivities may be computed efficiently using the adjoint method of sensitivity analysis. This can be done by introducing an intermediate vector variable, $\bsym{\Lambda}$, named the adjoint temperature, which is defined by the following:
\begin{equation}
    \label{eq: Lambda def 1}
    \mbf{M}\Tr \bsym{\Lambda} = N_t \left( \frac{\partial \Theta}{\partial \mbf{S}} \right)\Tr 
\end{equation}
which changes the sensitivity expression from \Cref{eq:sensitivity} to:
\begin{equation} \label{eq:sensitivity_final}
    \bsym{\nabla} \Theta = - \frac{1}{N_t} \bsym{\Lambda}\Tr \frac{\partial \mbf{R}}{ \partial \bsym{\chi} }
\end{equation}
The factor $N_t$ is included in the definition of $\bsym{\Lambda}$ for the purpose of making it so that the values of the entries of $\bsym{\Lambda}$ do not scale with $N_t$. The motivation for this will become apparent in \Cref{sec: choice of props}.

Solving for $\bsym{\Lambda}$ in \Cref{eq: Lambda def 1} is equivalent to solving the following terminal value problem:
\begin{subequations}
\begin{align}
    \bsym{\lambda}_{N_t+1} &= \mbf{0}
    \\
    \label{eq: euler method for lambda}
    \mbf{C}\Tr \frac{ \bsym{\lambda}_n - \bsym{\lambda}_{n+1} }{\Delta t} + \mbf{K}\Tr \bsym{\lambda}_n 
    &=
    N_t \left( \frac{\partial \Theta}{\partial \mbf{T}_n} \right)\Tr 
    \quad \text{for } n=N_t,N_t-1, \hdots 2,1
\end{align}
\end{subequations}
after which $\bsym{\Lambda}$ can be assembled as $\bsym{\Lambda} = \begin{pmatrix} \bsym{\lambda}_1\Tr & \bsym{\lambda}_2\Tr & \hdots & \bsym{\lambda}_{N_t}\Tr \end{pmatrix}\Tr$. The above terminal value problem is referred to as the adjoint problem. Note that this adjoint problem is very similar to the primal problem, since both $\mbf{C}$ and $\mbf{K}$ are symmetric. For the considered objective functional, the right-hand side of the above equation is the following:
\begin{align}
    N_t \left( \frac{\partial \Theta}{\partial \mbf{T}_n} \right)\Tr  
    &=
    \frac{\Theta^{1-p}}{N_e} {\mbf{T}_n}^{\circ (p-1)} 
\end{align}
where $ {\mbf{T}_n}^{\circ (p-1)} $ is the $(p-1)$\tss{th} Hadamard power of $\mbf{T}_n$, which is the vector $\mbf{T}_n$ after raising every element to the power $p-1$. Here it can be seen that the magnitude of the source term for the adjoint temperature does not depend on $N_t$, so $\bsym{\Lambda}$ should scale independently of $N_t$ as intended.

After computing $\bsym{\Lambda}$, the sensitivities can be evaluated as: 
\begin{align} \label{eq:sensitivities_sum}
    \bsym{\nabla} \Theta
    &=
    - \frac{1}{N_t} \sum_{n=1}^{N_t} \bsym{\lambda}_{n}\Tr \left( \frac{\partial \mbf{C} }{\partial \bsym{\chi}} \frac{\mbf{T}_n - \mbf{T}_{n-1}}{\Delta t} + \frac{\partial \mbf{K} }{\partial \bsym{\chi}} \mbf{T}_n \right) 
\end{align}
It can be seen that this formula for $\bsym{\nabla} \Theta$ depends on the values of the temperature at every time point. This observation will become important in \Cref{sec: modif parareal}.


\subsection{The Parareal algorithm}
\label{sec: parareal theory}


The Parareal algorithm is an iterative parallel-in-time method \cite{parareal_orig}, meaning that it is a method of parallelising the process of obtaining approximate solutions to IVPs. To explain how it works, the following general IVP is considered:
\begin{subequations}
\begin{align}
    \label{eq: parareal general IVP}
    \frac{\mathrm{d} \mathbf{u}(t) }{\mathrm{d} t} &= f(t, \mathbf{u}(t) )
    \\
    \mbf{u}(0) &= \mbf{u}_0
\end{align}
\end{subequations}
where $\mathbf{u}(t)$ is the vector function of time to be solved for, $f(t, \mathbf{u})$ is a given function, and $\mbf{u}_0$ is a given initial state. 

\begin{figure}[htbp]
\centering

\begin{tikzpicture}[scale=0.65]

\draw[black, thin, {}-{Stealth[scale=1.0]} ] (-0.6,0)  -- (15.6,0) node[anchor=west] {$t$}; 
\foreach \i in {0,...,15} \draw[thin, solid, black] (\i, -0.2) -- (\i, 0.2) node[anchor=south] {$t_{\i}$};
\foreach \i in {0,...,3} \draw[very thick, solid, black] (5*\i, 0.2) -- (5*\i, -0.5) node[anchor=north] {$\tau_{\i}$};

\end{tikzpicture}

\caption{Sketch of an example of how the time axis is coarsened when using the Parareal method for $N_t = 15$, $N_\tau = 3$, and $M = 5$. }
\label{fig: coarse time axis sketch}
\end{figure}
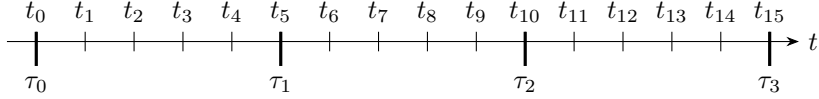

The output of the Parareal algorithm will not be evaluated on all of the time points $t_n$. Instead, it will be evaluated on a subset of these time points referred to as the coarse time points. These will be denoted $\tau_0, \tau_1, \hdots \tau_{N_\tau}$, where $N_{\tau}$ is the desired number of coarse time points (excluding the initial point). In this work, the coarse time points are defined to be:
\begin{align}
    \tau_n \equiv t_{M \cdot n} = M \cdot n \cdot \Delta t
\end{align}
where $M \equiv N_t / N_\tau$ is the coarsening factor. An example of such a set of coarse time points is sketched in \Cref{fig: coarse time axis sketch}. In this context, the notation $\mbf{u}_n$ will be used to refer to estimates of $\mbf{u}(\tau_n)$. Also note that $M$ must be an integer, which implies that $N_\tau$ must be a divisor of $N_t$. 

To get Parareal to work, it is necessary to define a time stepping method for it to use. This time-stepper is thought of as a function, denoted $\mathcal{F}(\tau_n, \mbf{u}_n)$, which is defined such that it treats $\mbf{u}(\tau_n) = \mbf{u}_n$ as being the initial condition and then returns a good estimate of $\mbf{u}(\tau_{n+1})$. The function $\mathcal{F}$ is referred to as the fine propagator. Parareal also needs a coarse propagator, denoted $\mathcal{G}(\tau_n, \mbf{u}_n)$. This should also return an estimate of $\mbf{u}(\tau_{n+1})$, but should be computationally cheaper to evaluate while being allowed to be less accurate. 

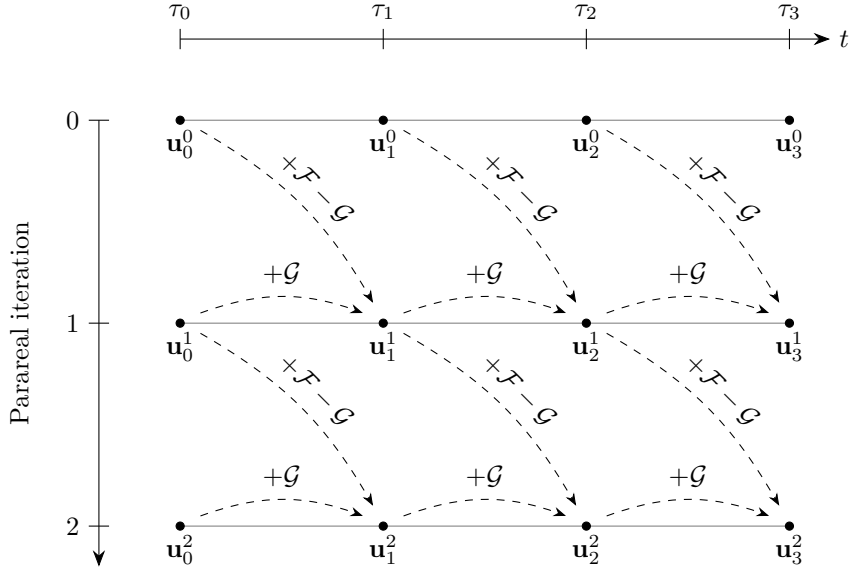
\begin{figure}[htbp]
\centering

\begin{tikzpicture}[scale=2.7]

\draw[black, thin, {}-{Stealth[scale=1.3]} ] (0,0.4)  -- (3.2,0.4) node[anchor=west] {$t$}; 
\foreach \n in {0,...,3} \draw[thin, solid, black] (\n, 0.35) -- (\n, 0.45) node[anchor=south] {$\tau_{\n}$};

\draw[black, thin, {}-{Stealth[scale=1.3]} ] (-0.4,0)  -- (-0.4,-2.2); 
\draw[black, thin] (-0.7,-1) node[anchor=south, rotate=90] {Parareal iteration}; 
\foreach \k in {0,...,2} \draw[thin, solid, black] (-0.35, -\k) -- (-0.45, -\k) node[anchor=east] {$\k$};

\foreach \k in {0,...,2} {
    
    \draw[gray, thin ] (0,-\k) -- (3,-\k); 

    \foreach \n in {0,...,3} {
        \filldraw[thin] (\n,-\k) circle (0.02) node[anchor=north] {$\mbf{u}_{\n}^{\k}$};
    }
}

\foreach \k in {0,...,1} { 
    \foreach \n in {0,...,2} {

        \draw[black, thin, dashed, {}-{Stealth[scale=1.0]} ] (\n+0.10,-\k-0.05) to[bend left=15] (\n+0.95,-\k-0.90);
        \draw[black] (\n+0.6,-\k-0.4) node[anchor=south, rotate=-45] {$\mathcal{+F - G}$};

        \draw[black, thin, dashed, {}-{Stealth[scale=1.0]} ] (\n+0.1,-\k-0.95)  to[bend left=20]  (\n+0.9,-\k-0.95);
        \draw[black] (\n+0.5,-\k-0.85) node[anchor=south] {$\mathcal{+G}$}; 
    }
}

\end{tikzpicture}

\caption{Diagram of the flow of information between different time points and Parareal iterations during execution of the Parareal algorithm, which is governed by \Cref{eq: parareal corr form orig}. Arrows labelled as $\mathcal{+G}$ indicate the addition of terms of the form $\mathcal{G}(\tau, \mbf{u})$ and arrows labelled as $\mathcal{+F-G}$ indicate the addition of terms of the form $\mathcal{F}(\tau, \mbf{u}) - \mathcal{G}(\tau, \mbf{u})$. }
\label{fig: parareal info flow diagram}
\end{figure}

The Parareal method is initialised by specifying an initial guess for the solution at every coarse time point, denoted $\mbf{u}_n^0$. The method then iteratively improves the solution using a correction step defined by the following recursive formula:
\begin{subequations}
\begin{align}
    \label{eq: parareal corr form orig}
    \mathbf{u}_{n+1}^{k+1} &= \mathcal{G}(\tau_n, \mathbf{u}_{n}^{k+1}) + \mathcal{F}(\tau_n, \mathbf{u}_{n}^{k}) - \mathcal{G}(\tau_n, \mathbf{u}_{n}^{k})
    \\
    \mbf{u}_0^k &= \mbf{u}_0
\end{align}
\end{subequations}
where $\mathbf{u}_{n}^{k}$ is the estimate of $\mbf{u}_n$ obtained at the $k$\tss{th} iteration of Parareal. This correction step is repeated until a specified convergence criterion is satisfied. \Cref{fig: parareal info flow diagram} shows a diagram of the flow of information associated with the above correction formula. As a whole, the correction formula has to be evaluated sequentially at each Parareal iteration. However, the terms involving $\mathcal{F}$ can be evaluated in parallel, thus allowing for speedup via parallelism.

Given enough iterations, Parareal will converge to the solution which would be obtained if the propagator $\mathcal{F}$ had been applied sequentially: $\mbf{u}_{n+1} = \mathcal{F}(\mbf{u}_{n})$. As such, when judging the error associated with Parareal, it makes sense to use this sequential solution as the reference solution. 

In this work, the fine and coarse propagators are distinguished from each other in the following way:
\begin{itemize}
    \item The propagator $\mathcal{F}$ is defined to compute $M$ time steps, each with a step size of $\Delta t$, going from $\tau_n$ to $\tau_{n+1}$. 
    \item The propagator $\mathcal{G}$ is defined to compute one time step with a step size of $\Delta\tau = t_T / N_\tau = M \cdot \Delta t$ going from $\tau_n$ to $\tau_{n+1}$. 
\end{itemize}
As such, the coarse propagator is computationally cheaper and less accurate than the fine propagator, as intended.


\subsection{Modified Parareal for accommodating adjoint sensitivity analysis}
\label{sec: modif parareal}


A problem presents itself when trying to compute the sensitivities, $\bsym{\nabla}\Theta$, using the output of Parareal. As noted in \Cref{sec: adj method}, it is necessary to know the temperature at all fine time points in order to compute the sensitivities using the adjoint method. However, Parareal only returns estimates for the temperature at the specific coarse time points $\tau_n$, not $t_n$. As a work-around for this problem, Parareal was modified to save the temperature field at each of the intermediate time points while evaluating the fine propagators, and return the saved fields as part of the solution at time points, where the correction formula of Parareal is not defined. 

\begin{figure}[phtb]
    \centering
    \includegraphics[width=0.9\linewidth]{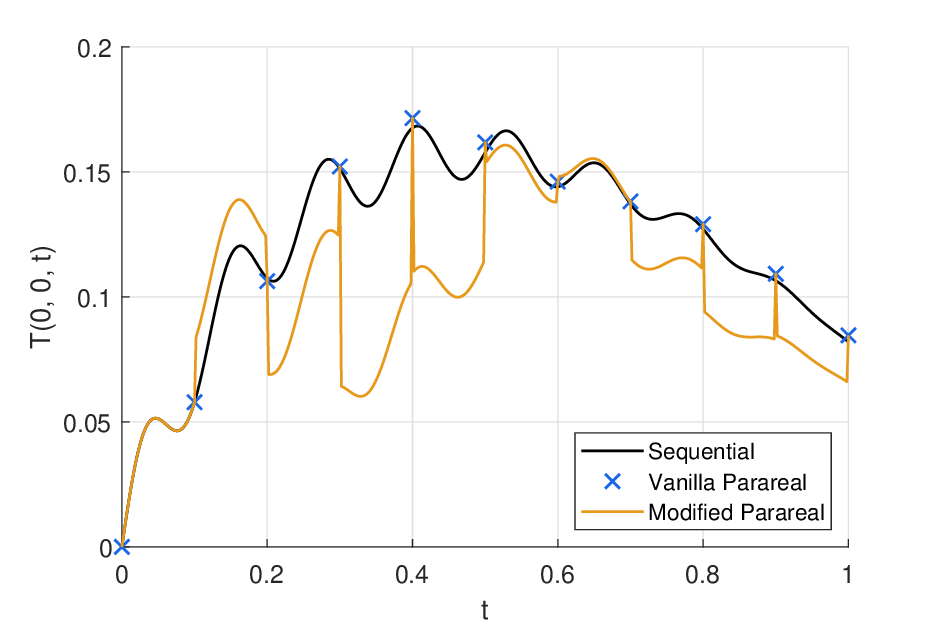}
    \caption{Examples of results returned by sequential time-stepping, the unmodified ``vanilla'' Parareal method, and the modified Parareal method (modified to save the states at the intermediate time steps). The plotted variable is the temperature at the centre of the domain as computed in the test described in \Cref{sec: prelim tests} where $N_\tau=10$ and one Parareal iteration was executed. }
    \label{fig: modif parareal accuracy demo}
\end{figure}

This modification can be interpreted as a variation of the MGRIT method, since Parareal is equivalent to some instances of two-level MGRIT, except for the fact that MGRIT also returns the solution obtained at the intermediate time steps \cite{mgrit_original}. Regardless of how it is interpreted, this modified implementation of Parareal is less efficient than unmodified ``vanilla" Parareal in terms of memory usage. It is also less efficient in terms of communication delays, because the current implementation of the modified Parareal is thread-based, so it spends time on sending the saved fields to the master thread. In addition, it returns results which are, on average, less accurate than unmodified Parareal, since unmodified Parareal only returns the parts of the solution where the correction step has been applied, while the modified Parareal method also returns parts where it has not been applied. An example of this is shown in \Cref{fig: modif parareal accuracy demo}, where it can clearly be seen that although the predicted solution at the coarse time points is decent, the solution at intermediate fine time points is quite poor.

\subsection{Choice of propagators used by Parareal}
\label{sec: choice of props}


Using the chosen coarsening approach, it is relatively straight-forward to define fine and coarse propagators which can be used by Parareal to solve for the temperature field. In addition, it is noted that it is possible to solve for the objective while solving for the temperature by using Parareal. This is done by defining a sequence of cumulative objective values, denoted $\theta_n$, at each time point $t_n$, by truncating the sum in the definition of the objective functional in \Cref{eq:objective_discrete} at the $n$\tss{th} term like so:
\begin{align}
    \theta_n &= \left( \frac{1}{N_t N_e} \sum_{j=1}^{n} {\lVert \mbf{T}_j \rVert_p}^p  \right)^{1/p} 
\end{align}
The total objective, $\Theta$, is then equal to $\theta_{N_t}$. Here it can be noted that $\theta_n$ can be expressed in terms of $\theta_{n-1}$ and $\mbf{T}_n$ like so:
\begin{align}
    \theta_n &= \left( {\theta_{n-1}}^p + \dfrac{1}{N_t N_e} {\lVert \mbf{T}_n \rVert_p}^p \right)^{1/p}
\end{align}
This lends itself to defining a propagator for the cumulative objective which is coupled to the temperature. Based on this consideration, the vectors which are solved for by Parareal are defined to be of the following form:
\begin{align}
    \mbf{u}_n 
    &= 
    \begin{pmatrix}
        \mbf{T}_{M \cdot n} \\ \theta_{M \cdot n}
    \end{pmatrix}
\end{align}
The variables on the right-hand side have subscripts $M \cdot n$ because $\mbf{u}$ is defined only on the coarse time points, while $\mbf{T}$ and $\theta$ are defined on all time points. The initial condition, $\mbf{u}_0$, (not to be confused with the initial guess, $\mbf{u}_n^0$) is $\mbf{0}$, since both $\mbf{T}_0 = \mbf{0}$ and $\theta_0 = 0$.

To help define the propagators used by Parareal, a propagator denoted $\mathcal{H}_{pri}$ is defined in \Cref{alg: primal single step prop} to compute one step of the backward Euler method for the temperature and one step of accumulating the objective.
\begin{algorithm}
\caption{Definition of propagator $\mathcal{H}_{pri}$}
\label{alg: primal single step prop}
\begin{algorithmic}[1]
\ENSURE $
\begin{pmatrix} \mbf{T}_{n} \\ \theta_{n} \end{pmatrix}
=
\mathcal{H}_{pri} \left[ t_{n-1}, \begin{pmatrix} \mbf{T}_{n-1} \\ \theta_{n-1} \end{pmatrix} \right] 
$ 

\STATE Solve for $\mbf{T}_{n}$ in the equation $\mbf{C} \dfrac{\mbf{T}_n - \mbf{T}_{n-1}}{\Delta t} + \mbf{K} \mbf{T}_n = \mbf{q}_n$

\STATE $\theta_n \gets \left( {\theta_{n-1}}^p + \dfrac{1}{N_t N_e} {\lVert \mbf{T}_n \rVert_p}^p \right)^{1/p}$

\end{algorithmic}
\end{algorithm}
The fine propagator for the primal problem, $\mathcal{F}_{pri}$, is then defined to repeat $\mathcal{H}_{pri}$ $M$ times, while the coarse propagator for the primal problem, $\mathcal{G}_{pri}$, is defined to be the same as $\mathcal{H}_{pri}$, but with the following substitutions:
\begin{itemize}
    \item $\Delta t$ should be replaced with $\Delta \tau$,
    \item $N_t$ should be replaced with $N_\tau$, 
    \item and every variable with subscript $n-1$ should instead have subscript $n-M$.  
\end{itemize}

The adjoint problem is also solved using Parareal. For this purpose, the direction of time for the correction formula of Parareal must be reversed, since the adjoint problem is a terminal value problem. As such, the correction formula becomes:
\begin{align}
    \mathbf{v}_{n-1}^{k+1} &= \mathcal{G}(\tau_n, \mathbf{v}_{n}^{k+1}) + \mathcal{F}(\tau_n, \mathbf{v}_{n}^{k}) - \mathcal{G}(\tau_n, \mathbf{v}_{n}^{k})
\end{align}
Here $\mbf{v}$ is used to denote the state vectors of the adjoint problem in order to distinguish them from those of the primal problem. 

Similarly to how the temperature and objective are solved simultaneously using Parareal, the adjoint temperature and the sensitivities are also solved for simultaneously using Parareal. This is done by introducing cumulative sensitivities at each time point by truncating the sum in \Cref{eq:sensitivities_sum}:
\begin{align}
    \mbf{g}_n
    &=
    - \frac{1}{N_t} \sum_{j=n}^{N_t} \bsym{\lambda}_{j}\Tr \left( \frac{\partial \mbf{C} }{\partial \bsym{\chi}} \frac{\mbf{T}_j - \mbf{T}_{j-1}}{\Delta t} + \frac{\partial \mbf{K} }{\partial \bsym{\chi}} \mbf{T}_j \right) 
\end{align}
Note that this sum is truncated from below instead of from above. As such, the total gradient is $\bsym{\nabla} \Theta = \mbf{g}_1$. It also makes it so that $\mbf{g}_n$ can be expressed in terms of $\mbf{g}_{n+1}$ and $\bsym{\lambda}_{n}$ as:
\begin{align}
    \mbf{g}_n 
    &=
    \mbf{g}_{n+1} - \frac{1}{N_t} \bsym{\lambda}_{n}\Tr \left( \dfrac{\partial \mbf{C} }{\partial \bsym{\chi}} \dfrac{\mbf{T}_n - \mbf{T}_{n-1}}{\Delta t} + \dfrac{\partial \mbf{K} }{\partial \bsym{\chi}} \mbf{T}_n \right) 
\end{align}
This lends itself to defining backward-in-time propagators for $\bsym{\lambda}$ and $\mbf{g}$. As such, the vectors solved for by Parareal for the adjoint problem are defined to be of the following form:
\begin{align}
    \mbf{v}_n = \begin{pmatrix} \bsym{\lambda}_{M \cdot n + 1} \\ \mbf{g}_{M \cdot n + 1}\Tr \end{pmatrix}
\end{align}
The $+1$ is added in the subscripts to make it so that $\mbf{v}_0$ contains $\mbf{g}_1$. The terminal condition for this vector is $\mbf{v}_{N_\tau} = \mbf{0}$. Similarly to the primal problem, a single step propagator, denoted $\mathcal{H}_{adj}$, is defined for the adjoint problem in \Cref{alg: adjoint single step prop}. 
\begin{algorithm}
\caption{Definition of propagator $\mathcal{H}_{adj}$}
\label{alg: adjoint single step prop}
\begin{algorithmic}[1]
\ENSURE $
\begin{pmatrix} \bsym{\lambda}_{n} \\ \mbf{g}_{n}\Tr \end{pmatrix}
=
\mathcal{H}_{adj} \left[ t_{n+1}, \begin{pmatrix} \bsym{\lambda}_{n+1} \\ \mbf{g}_{n+1}\Tr \end{pmatrix} \right] 
$

\STATE Solve for $\bsym{\lambda}_n$ in the equation $\mbf{C}\Tr \dfrac{\bsym{\lambda}_n - \bsym{\lambda}_{n+1}}{\Delta t} + \mbf{K}\Tr \bsym{\lambda}_n = \dfrac{\Theta^{1-p}}{N_e} {\mbf{T}_n}^{\circ (p-1)}  $

\STATE $\mbf{g}_n \gets \mbf{g}_{n+1} - \dfrac{1}{N_t} \bsym{\lambda}_{n}\Tr \left( \dfrac{\partial \mbf{C} }{\partial \bsym{\chi}} \dfrac{\mbf{T}_n - \mbf{T}_{n-1}}{\Delta t} + \dfrac{\partial \mbf{K} }{\partial \bsym{\chi}} \mbf{T}_n \right) $

\end{algorithmic}
\end{algorithm}
The fine propagator for the adjoint problem, $\mathcal{F}_{adj}$, is then defined to repeat $\mathcal{H}_{adj}$ $M$ times, while the coarse propagator for the primal problem, $\mathcal{G}_{adj}$, is defined to be the same as $\mathcal{H}_{adj}$, but with the following substitutions:
\begin{itemize}
    \item $\Delta t$ should be replaced with $\Delta \tau$,
    \item $N_t$ should be replaced with $N_\tau$, 
    \item and every variable with subscript $n \pm 1$ should instead have subscript $n \pm M$. 
\end{itemize}

Here it should be recalled that $\bsym{\Lambda}$ (and subsequently $\bsym{\lambda}$) was defined in \Cref{sec: adj method} in such a way, that its entries do not scale with $N_t$. The motivation for this lack of scaling is to make the output of $\mathcal{G}_{adj}$ scale the same way as the output of $\mathcal{F}_{adj}$, since the fine and coarse propagators used by Parareal are supposed to return approximate solutions of the same problem. 
Unmodified Parareal is used to solve the adjoint problem instead of the modified Parareal method, because there is no need to store the intermediate values of $\bsym{\lambda}$ and $\mbf{g}$ for later. This cuts down on memory usage and communication delays.

In summary, the temperature, objective, adjoint temperature, and sensitivities will be estimated using Parareal by doing the following: 
\begin{enumerate}
    \item Solve for the temperatures, $\mbf{T}$, and cumulative objective values, $\theta$, using the version of Parareal which was modified to save the output at the intermediate time points.
    \item Extract the total objective by setting $\Theta = \theta_{N_t}$.
    \item Solve for the adjoint temperatures, $\bsym{\lambda}$, and cumulative sensitivities, $\mbf{g}$, using vanilla Parareal, except going backwards in time. 
    \item Extract the total sensitivities by setting $\bsym{\nabla} \Theta = \mbf{g}_1$. 
\end{enumerate}


\section{Preliminary tests of Parareal}
\label{sec: prelim tests}


To test the performance of the proposed method, some preliminary tests are made on a fixed design field wherein the speed and accuracy of the method is measured. 


\subsection{Definition of test case}
\label{sec: test case def}

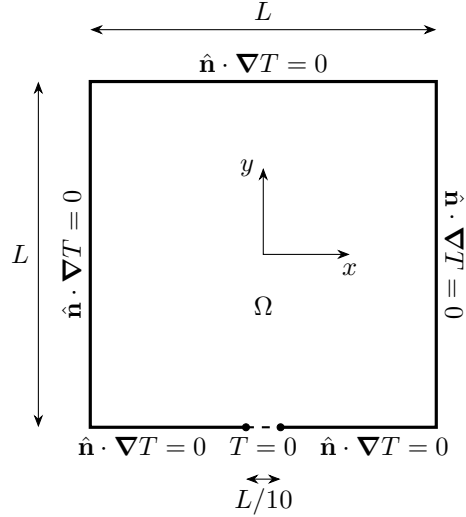
\begin{figure}[phtb]
\centering
\begin{tikzpicture}[scale=0.23]

\draw[draw=black, thin, {Stealth[scale=1.0]}-{Stealth[scale=1.0]} ] (0,5) node[anchor=east]{$y$} -- (0,0) -- (5,0) node[anchor=north]{$x$};

\draw[thick, dashed, black] (-1,-10) -- (1,-10);
\draw[very thick, solid] (-1,-10) -- (-10,-10) -- (-10,10) -- (10,10) -- (10,-10) -- (1, -10); 

\filldraw [fill=black, draw=black] (-1,-10) circle (0.2); 
\filldraw [fill=black, draw=black] ( 1,-10) circle (0.2); 

\draw[black, thin, {Stealth[scale=1.0]}-{Stealth[scale=1.0]} ] (-10,13)  -- (0,13) node[anchor=south]{$L$} -- (10,13); 
\draw[black, thin, {Stealth[scale=1.0]}-{Stealth[scale=1.0]} ] (-13,10)  -- (-13,0) node[anchor=east]{$L$} -- (-13,-10); 
\draw[black, thin, {Stealth[scale=1.0]}-{Stealth[scale=1.0]} ] (-1,-13)  -- (0,-13) node[anchor=north]{$L/10$} -- (1,-13); 

\draw [black] (0, -10) node[anchor=north, align=center]{$ T=0 $}; 
\draw [black] (-10, 0) node[anchor=south, rotate=90, align=center]{$ \hat{\mbf{n}} \cdot \bsym{\nabla} T = 0 $}; 
\draw [black] (0, 10) node[anchor=south, align=center]{$ \hat{\mbf{n}} \cdot \bsym{\nabla} T = 0 $}; 
\draw [black] (10, 0) node[anchor=south, rotate=-90, align=center]{$ \hat{\mbf{n}} \cdot \bsym{\nabla} T = 0 $}; 
\draw [black] (-7, -10) node[anchor=north, align=center]{$ \hat{\mbf{n}} \cdot \bsym{\nabla} T = 0 $}; 
\draw [black] ( 7, -10) node[anchor=north, align=center]{$ \hat{\mbf{n}} \cdot \bsym{\nabla} T = 0 $}; 

\draw [black] (0, -3) node[anchor=center, align=center]{$ \Omega $};

\end{tikzpicture}
\caption{Geometry and boundary conditions of the domain of the test case defined in \Cref{sec: test case def}. }
\label{fig: geometry sketch}
\end{figure}

The method was tested on a system consisting of a square domain with a side length of $L$. The boundary conditions of the system are sketched in \Cref{fig: geometry sketch} and consist of homogeneous Neumann conditions on all boundaries except for a small part at the bottom with a homogeneous Dirichlet condition. In addition, the depth of the domain (in the out-of-plane direction) is also set equal to $L$.

\begin{figure}[phtb]
    \centering
    \includegraphics[width=0.8\linewidth]{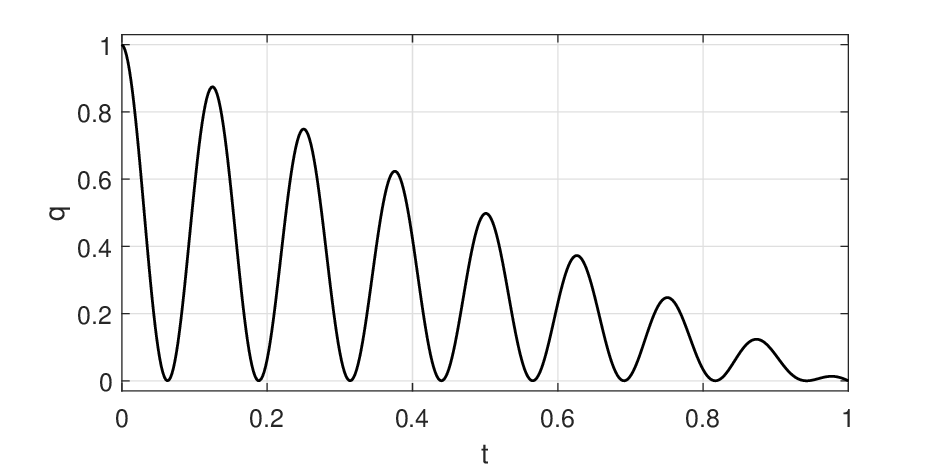}
    \caption{Imposed heat load, $q$ as a function of time, $t$, as defined in \Cref{eq: imposed heat for test}. }
    \label{fig: q over t plot}
\end{figure}

The system is non-dimensionalised by setting $L=1$, $t_T=1$, and $c_0=1$. A complete list of the chosen material parameters and penalty parameters can be found in \Cref{tab: mat pen fil params}. The imposed heat load is defined to be the following:
\begin{align}
    \label{eq: imposed heat for test}
    q(\mbf{x}, t) &= \frac{1}{2} (1-t) \left[ 1 + \cos(50t) \right]  
\end{align} 
This external heat load is uniform over space but variable over time, and it oscillates with a constant frequency and variable amplitude, as shown in \Cref{fig: q over t plot}. The frequency is high enough that it makes it difficult for the coarse propagators to capture the oscillation, at least for small values of $N_\tau$. The domain is discretised with a uniform mesh of $100 \times 100$ square elements and the total number of time points is set to $N_t = 480$. 

\begin{table}[phtb]
\centering
\begin{tabular}{|l|llllll|}
\hline
\textbf{Parameter:} &
$k_0$ & $k_{min}$ & $c_0$ & $c_{min}$ & $p_k$ & $p_c$  \\ 
\textbf{Value:} &
$3$   & $0.03$    & $1$   & $0.5$     & $3$   & $2$ \\ 
\hline
\end{tabular}
\caption{List of material parameters and penalty parameters chosen for the test case defined in \Cref{sec: test case def}. }
\label{tab: mat pen fil params}
\end{table}

\subsection{Procedure for preliminary tests}
\label{sec: prelim test procedure}


The purpose of the preliminary tests is the following:
\begin{itemize}
    \item To evaluate how accurately the proposed Parareal-based method can compute the objective and sensitivities (since these are the pieces of information which have to be passed to MMA); 
    \item To measure how much speedup the Parareal-based method provides. 
\end{itemize}
The aim is not to evaluate the error associated with the finite element method and backward Euler method. As such, the reference method, which the solutions and wall-clock times will be compared against, is one where $\mathcal{F}_{pri}$ and $\mathcal{F}_{adj}$ are applied sequentially, as advocated for in \Cref{sec: parareal theory} (or rather, it is $\mathcal{H}_{pri}$ that is applied sequentially when solving the primal problem, since the temperature at every time point has to be stored). 

The errors associated with the objective and sensitivities will be evaluated like so:
\begin{align}
    \text{Error in objective}
    &= \frac{ |\Theta_{par} - \Theta_{seq}| }{ |\Theta_{seq}| } 
    \\ 
    \text{Error in sensitivity}
    &= \frac{ \lVert \bsym{\nabla} \Theta_{par} - \bsym{\nabla} \Theta_{seq} \rVert_2 }{ \lVert \bsym{\nabla} \Theta_{seq} \rVert_2 } 
\end{align}
where $\Theta_{par}$ is the objective estimated by Parareal and $\Theta_{seq}$ is the true objective evaluated using the sequential reference method. Similarly, $\bsym{\nabla} \Theta_{par}$ and $\bsym{\nabla} \Theta_{seq}$ are sensitivities estimated by Parareal and the reference method respectively, and $\lVert \cdot \rVert_2$ is the Euclidean norm. 

For these preliminary tests, the design field was fixed to a topology obtained after performing topology optimisation on the defined test case by using a method which exclusively used sequential time-stepping. 

The initial guess solutions supplied to Parareal are generated using the coarse propagators, such that $\mbf{u}_{n+1}^0 = \mathcal{G}_{pri}(\tau_n, \mbf{u}_{n}^0)$ and $\mbf{v}_{n-1}^0 = \mathcal{G}_{adj}(\tau_n, \mbf{v}_{n}^0)$. 

The algorithm is implemented in MATLAB and parallelised using parallel pools of worker threads facilitated by MATLAB's Parallel Computing Toolbox. In each test, the number of worker threads in the pool is set equal to $N_\tau$. Four values of $N_\tau$ are considered for the preliminary tests: $N_\tau = 5$, 10, 20, and 30. For each value of $N_\tau$, the number of applied Parareal iterations is varied between 1 and 8. The tests are executed on a machine consisting of two Intel Xeon Gold 6130 processors, meaning the machine had a total of 32 cores and 64 threads. 

During early tests of the code, technical difficulties were encountered when measuring the wall-clock time of the sequential time-stepping code. For more information about these technical difficulties and how they were mitigated during the subsequent tests, see \Cref{sec: timing quirks}.


\subsection{Results of preliminary tests}
\label{sec: prelim test results}

\begin{figure}[htb]
    \centering
    \includegraphics[width=0.95\textwidth]{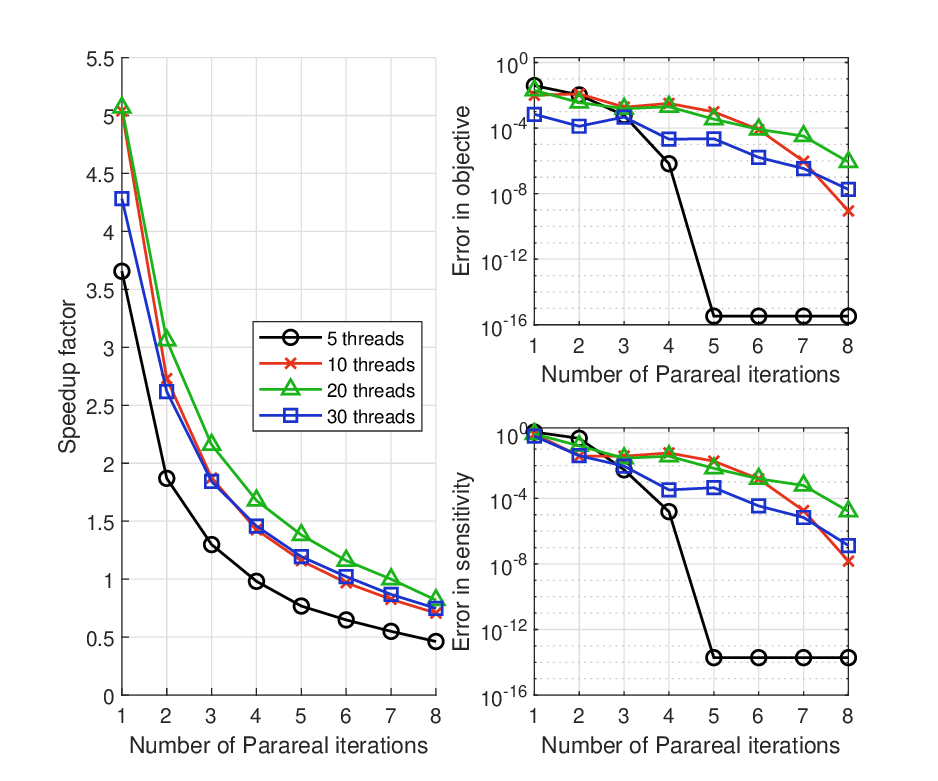}
    \caption{Plots of the results of the preliminary tests described in \Cref{sec: prelim tests}.}
    \label{fig: prelim test all results}
\end{figure}

The results of the preliminary tests are plotted in \Cref{fig: prelim test all results}. It can be seen that the parallel efficiency is not particularly impressive. The speedup gained when using 5 threads never goes above $4\times$, and the speedup gained with 10 - 30 threads only goes slightly above $5\times$. 
It can also be seen that the errors do not decrease monotonically as the number of Parareal iterations goes up. Additionally, the errors in the sensitivities are generally larger than the errors in the objective. This makes sense, because the sensitivities depend on the temperatures and the total objective, which means that the errors in the temperatures and objectives will contribute to the errors in the sensitivities. 

If a fairly loose tolerance of $10^{-2}$ is imposed on the errors in the sensitivities, then the best measured speedup that satisfies this tolerance is $1.8 \times$, which is at 3 iterations using 30 threads. If the tolerance is lowered to $10^{-3}$, then the best measured speedup drops to $1.5\times$ at 4 iterations using 30 threads. These speedup factors are unimpressive and borderline useless. Thus, it would be convenient to have a method which combines the speedup achieved for one iteration while also having the accuracy achieved at greater numbers of iterations. This is where one-shot methods come in, which will be explained in the next section.


\section{One-shot Parareal method of topology optimisation}
\label{sec: one-shot approach}


To maximise the speedup gained from the use of Parareal, while still having acceptable accuracy, a ``one-shot" approach is applied to the considered topology optimisation problem. In this context, being a one-shot method means that an iterative method (Parareal) is used as part of the optimisation process, and for each optimisation cycle, only one iteration of this iterative method is performed to solve the primal problem and only one iteration is performed to solve the adjoint problem. In addition, the initial guesses provided to Parareal are set to be the primal/adjoint solution obtained at the previous iteration. This way of setting the initial guess is referred to as a ``warm restart". Of course, this does not work at the first iteration of the optimisation loop. Instead, the fine propagators for the primal and adjoint problems are applied sequentially to obtain the solutions at the first iteration. The proposed one-shot approach is presented in the form of pseudocode in \Cref{alg: one-shot}.

\begin{algorithm}
\caption{One-shot Parareal Optimisation Method}
\label{alg: one-shot}
\begin{algorithmic}[1]
\STATE Initialise system parameters, MMA, etc. 
\STATE $i \gets 0$
\WHILE{$i < i_{max}$ }
    \STATE $i \gets i + 1$
    \IF{$i = 1$}
        \STATE Solve for $\mbf{S}$ and $\theta$ using sequential time-stepping. 
        \STATE Solve for $\bsym{\Lambda}$ and $\mbf{g}$ using sequential time-stepping. 
    \ELSE
        \STATE \label{line: one-shot parareal primal} Solve for $\mbf{S}$ and $\theta$ using 1 iteration of Parareal with guess being the solution obtained at the previous iteration. 
        \STATE \label{line: one-shot parareal adj} Solve for $\bsym{\Lambda}$ and $\mbf{g}$ using 1 iteration of Parareal with guess being the solution obtained at the previous iteration. 
    \ENDIF
    \STATE $\Theta \gets \theta_{N_t}$
    \STATE $\bsym{\nabla} \Theta \gets \mbf{g}_1$
    \STATE Call MMA subroutine to update $\bsym{\chi}$ using the obtained $\bsym{\nabla}\Theta$ and $\Theta$. 
\ENDWHILE
\end{algorithmic}
\end{algorithm}

The idea behind this one-shot approach is that if the design field only changes by small amounts between each design update, then the solutions for both the primal and adjoint problems are expected to be similar to the solutions obtained at the previous optimisation iteration. Therefore, it should be possible to obtain good solutions at each iteration simply by applying a small correction to the solutions obtained at the previous iteration. This rule typically applies well to the later iterations of the optimisation process, where the design changes quite slowly. However, for the earlier iterations, the design field typically changes quite rapidly, thus making the rule less applicable, but the one-shot approach is performed nonetheless. 


\subsection{Procedure for tests of one-shot Parareal method}
\label{sec: one-shot parareal exp procedure}


To test the proposed one-shot method of topology optimisation, it is implemented in MATLAB and parallelised using parallel pools of worker threads facilitated by MATLAB's Parallel Computing Toolbox. It is applied to the test case defined in \Cref{sec: test case def}. 

The value of $N_\tau$ is varied between each test. Specifically, $N_\tau$ is set to assume the value of every divisor of 480 between 2 and 32, which means that 14 different tests are executed (or 15 if the test of the sequential reference method is included). The number of worker threads in the parallel pool is always set equal to $N_\tau$. The reference method is one that is identical to the proposed one-shot method, except it uses sequential time-stepping at every iteration by applying the fine propagators sequentially. When measuring the wall-clock time of this reference method, the method of timing described in \Cref{sec: timing quirks} is used. 

The chosen values for the parameters for the threshold projection filter are $\beta=32$, $\eta=0.5$, and $r_{fil} = 0.03$. The maximum area fraction is set to $a_{max} = 0.3$ and the initial design field is set to be uniform such that $\chi_{phys}(\mbf{x}) = a_{max}$ at every point in the domain. After that, 300 optimisation iterations are executed. During this, no continuation scheme is applied. 

For the sake of accurately comparing the final topologies obtained in each test, the true objective (as opposed to the objective estimated by Parareal) is evaluated at the final iteration of each test. In addition, for the sake of accurately judging the rate of convergence of the one-shot method, the true objective is computed and stored at every iteration in the specific tests where $N_\tau = 5$, 10, and 30. For these tests, the time taken to compute the true objective is excluded from the wall-clock time when the speedup is evaluated.


\subsection{Results of tests of one-shot Parareal method}
\label{sec: one-shot parareal results}

\begin{figure}[phtb]
    \centering
    \includegraphics[width=0.8\linewidth]{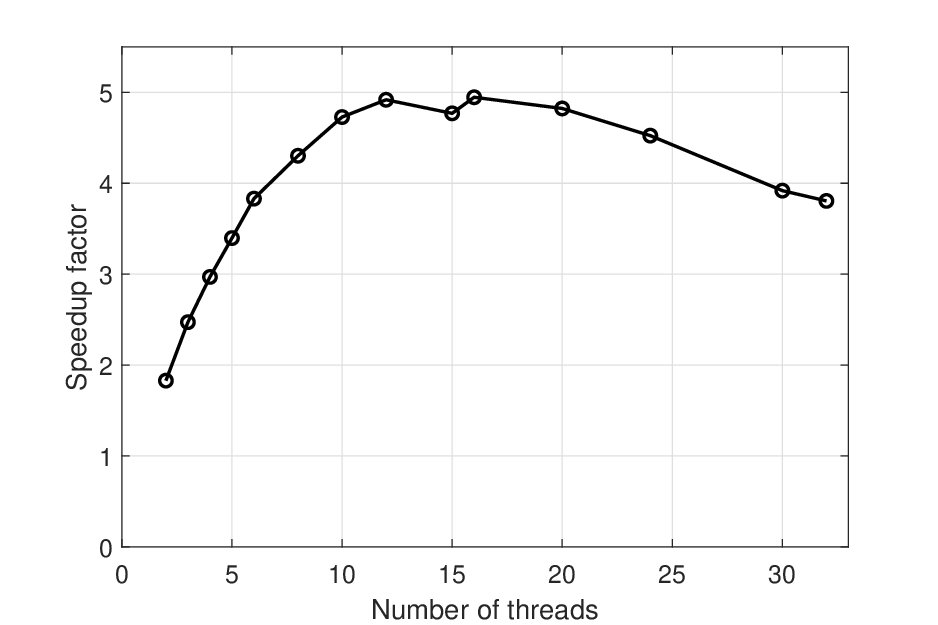}
    \caption{Measured speedup of the one-shot Parareal method of topology optimisation. The peak is at 16 threads with a speedup of $ 4.95 \times $. }
    \label{fig: one-shot parareal speedup}
\end{figure}

The measured speedup factors of each test are shown in \Cref{fig: one-shot parareal speedup}. The peak is at a speedup of $4.95\times$, which is consistent with the speedup factors observed in \Cref{sec: prelim test results}. 

\begin{figure}[phtb]
    \centering
    \includegraphics[width=0.9\linewidth]{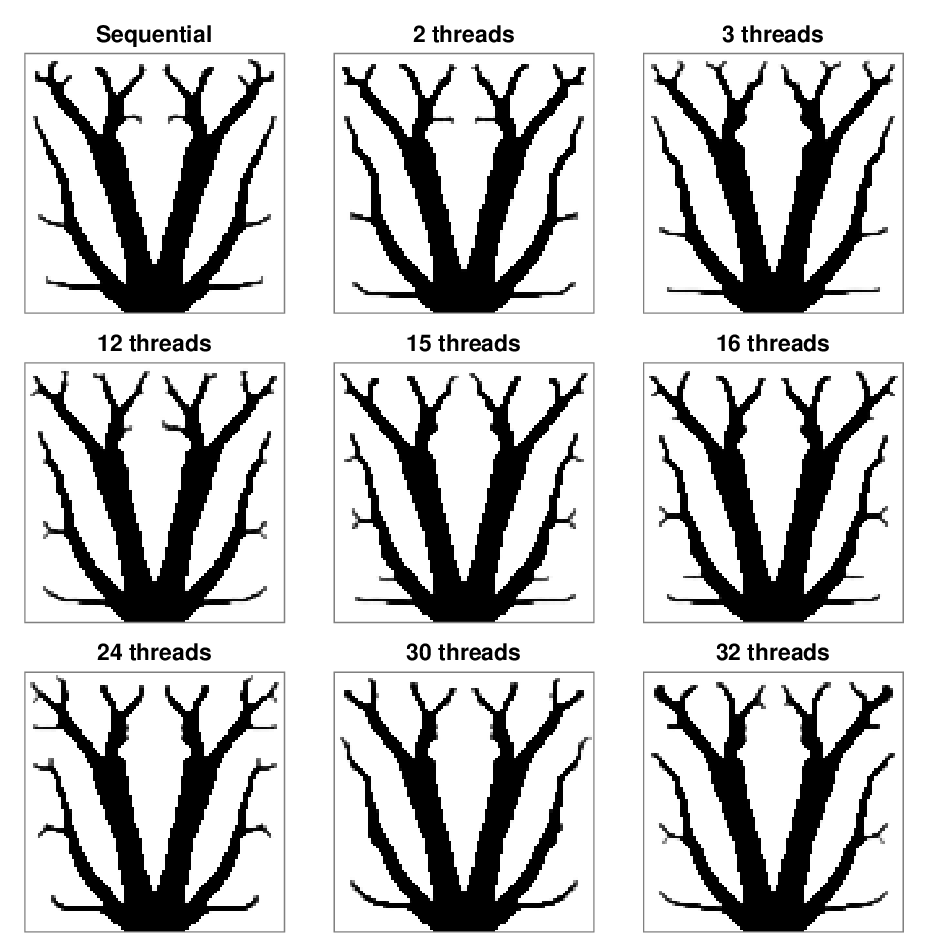}
    \caption{A selection of nine different local minima of the objective function that were found using the sequential reference method and the one-shot Parareal method with different numbers of threads. }
    \label{fig: one-shot parareal nine designs}
\end{figure}

The tests converged to different local minima of the objective due to the fact that the approximations of $\Theta$ and $\bsym{\nabla}\Theta$ depend on $N_\tau$, which in turn is equal to the number of threads. \Cref{fig: one-shot parareal nine designs} shows a selection of eight of the designs that were obtained at the last iteration of tests of the one-shot Parareal method, along with the reference method. It can be seen that the resulting designs look qualitatively similar, but they do have different placements of small branches. 
It can also be seen that the designs are not perfectly symmetrical (especially noticeable in the case with 12 threads) despite the fact that the system and the initial design field are both symmetrical.
The asymmetry is most likely ``seeded" by small inaccuracies in the solvers, after which the amount of asymmetry is amplified by the MMA subroutine. The underlying optimisation problem is highly non-convex and multimodal, so it is easy to end up in a different local minima given the slight inaccuracies. There are also examples of non-symmetric solutions being optimal for symmetric problems \cite{Rozvany2011}.

\begin{figure}[phtb]
    \centering
    \includegraphics[width=0.8\linewidth]{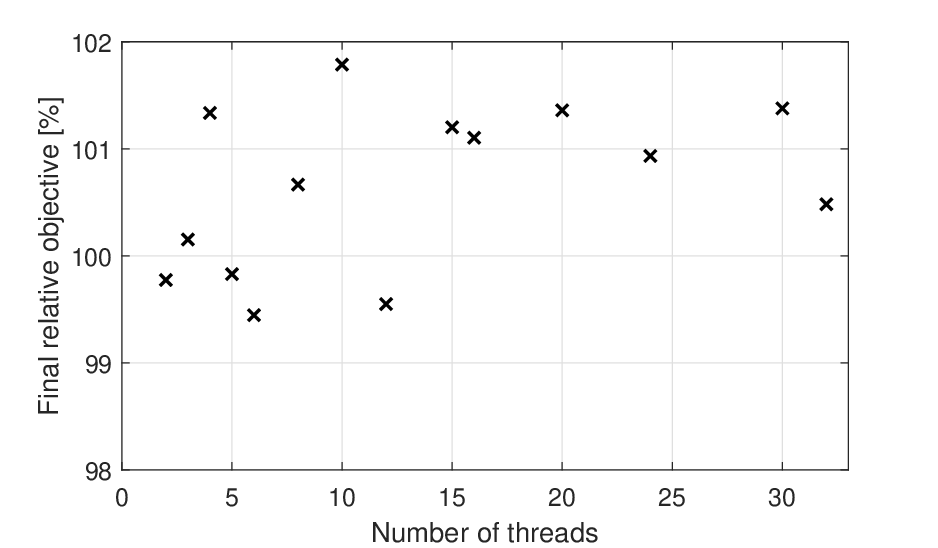}
    \caption{Relative objective values of the designs obtained at the final iterations of the tests of the one-shot Parareal method of topology optimisation. These are shown as a percentage of the final objective obtained by the sequential reference method. }
    \label{fig: one-shot parareal final obj}
\end{figure}

\begin{figure}[phtb]
    \centering
    \includegraphics[width=0.8\linewidth]{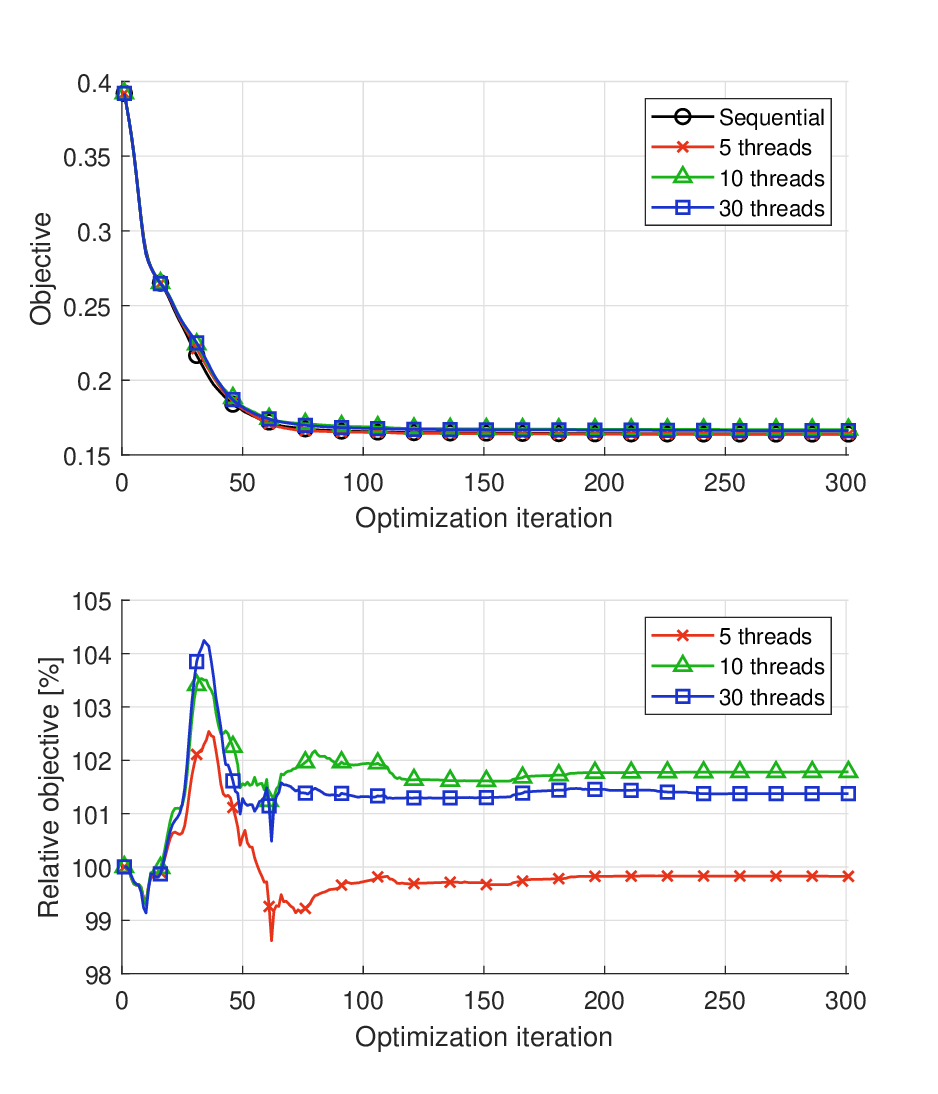}
    \caption{\textbf{Top:} True objective values (as opposed to objective values estimated by Parareal) as a function of the optimisation iteration during a small selection of the tests of the one-shot Parareal method, along with the test of the sequential reference method. \textbf{Bottom:} The relative values of the objective histories shown at the top, normalised with respect to the objective values of the reference method at each iteration. }
    \label{fig: one-shot parareal obj hist}
\end{figure}

\Cref{fig: one-shot parareal final obj} shows the relative final objective values found in each test, given as a percentage of the final objective found by the reference method. It can be seen that the objective values of the final designs are all within $\pm 2 \%$ of the objective obtained by the reference method. In particular, there are even a handful of designs (where $N_\tau = 2$, 5, 6, and 12) which are slightly better than the design found by the reference method. 
The true objective for the tests where $N_\tau = 5$, 10, and 30, along with the test of the reference method, are plotted as functions of the optimisation iteration at the top of \Cref{fig: one-shot parareal obj hist}. These objective histories are almost indistinguishable, so at the bottom of \Cref{fig: one-shot parareal obj hist} they are shown relative to the objective history of the reference method. Here it can be seen that the one-shot Parareal method generally improves the objective more slowly than the reference method during the first 50 iterations, as it lags up to 4\% behind the reference method. However, after the 50\tss{th} iteration, the one-shot Parareal method catches up with the reference method to some extent. In the test where $N_\tau = 5$, the one-shot method even overtakes the reference method. After about 100 iterations, both methods become more or less stagnant. 

Based on the above observations, it is seen that the one-shot Parareal method of topology optimisation is capable of finding designs comparable to the sequential reference method, both in terms of qualitative appearance and objective values. However, if the one-shot Parareal method is terminated prematurely, then it returns results which are consistently worse than what the reference method would find if it were terminated at the same number of iterations. This makes sense, considering that the estimates of the sensitivities are expected to be worse during the earlier iterations of the one-shot method. However, since the one-shot Parareal method takes less wall-clock time per optimisation cycle, it is likely that the one-shot Parareal method will return better results than the reference method, if both are stopped prematurely after the same amount of real time.


\section{Comparison to the Parallel Local-in-Time method of topology optimisation}
\label{sec: PLT comparison}


Recall that the coarse propagators used by Parareal are intended to have two properties: they should be cheap to evaluate, while being allowed to be inaccurate. These two properties can be taken to their extreme by making the coarse propagators not perform any processing at all, thus returning results which are completely inaccurate. For example, the coarse propagators could be set to only return the zero vector:
\begin{align}
    \mathcal{G}_{pri}(\tau_n, \mbf{u}_n) = \mbf{0}, \quad \mathcal{G}_{adj}(\tau_n, \mbf{v}_n) = \mbf{0}
\end{align}
In this case, the correction formulae used by Parareal simply become the following:
\begin{subequations}
\begin{align}
    \mbf{u}_{n+1}^{k+1} &= \mathcal{F}_{pri}(\tau_n, \mbf{u}_{n}^{k}) 
    \\
    \mbf{v}_{n-1}^{k+1} &= \mathcal{F}_{adj}(\tau_n, \mbf{v}_{n}^{k}) 
\end{align}
\end{subequations}
If this were to be plugged into the one-shot Parareal method proposed in this paper, then the result would be a method which is very similar to the ``Parallel Local-in-Time'' (PLT) method of topology optimisation \cite{Theulings_PLT}. In the PLT method, $\mbf{u}$ and $\mbf{v}$ at the $(k+1)$\tss{th} optimisation iteration are evaluated according the two formulae above, using the values found at the $k$\tss{th} iteration. However, the vectors $\mbf{u}$ and $\mbf{v}$ in the PLT method only contain the state vectors and adjoint state vectors, unlike the method proposed in this paper where they also contain cumulative objectives and cumulative sensitivities. Instead, the PLT method can be thought of as evaluating objectives and sensitivities in post-processing at each optimisation iteration. For more information on the PLT method, refer to \cite{Theulings_PLT}. Based on these considerations, it can be said that the PLT method can be interpreted as a one-shot method which uses Parareal where the coarse propagators have been set to zero.


\subsection{Procedure for tests of Parallel Local-in-Time method}


The PLT method described in \cite{Theulings_PLT} assumes that the objective is of the following form:
\begin{align}
    \Theta &= \sum_{j=0}^{N_t} F_j( \mbf{T}_j, \bsym{\chi} ) \Delta t
\end{align}
where $F_j$ is a sequence of given functions. This is a problem, because the objective function considered in this paper is not of this form. As a work-around for this, the following modified objective is introduced:
\begin{equation}
    \widetilde{\Theta} 
    = 
    \Theta^p 
    = 
    \sum_{j=1}^{N_t} \frac{1}{N_t N_e} {\lVert \mbf{T}_j \rVert_p}^p =
    \sum_{j=1}^{N_t} \frac{1}{t_T N_e} {\lVert \mbf{T}_j \rVert_p}^p \, \Delta t
\end{equation}
The adjoint sensitivity analysis for this modified objective is identical to that of the original objective, except the source term for the adjoint temperature is:
\begin{align}
    N_t \left( \frac{\partial \widetilde{\Theta}}{\partial \mbf{T}_n} \right)\Tr  
    &=
    \frac{p}{N_e} {\mbf{T}_n}^{\circ (p-1)} 
\end{align}
After computing $\widetilde{\Theta}$ and $\bsym{\nabla} \widetilde{\Theta}$, the original objective and gradient can be obtained using the following formulae: 
\begin{align}
    \label{eq: obj from modded obj}
    \Theta &= \widetilde{\Theta}^{1/p}
    \\
    \label{eq: nabla obj from modded nabla obj}
    \bsym{\nabla} \Theta &= \frac{\Theta^{1-p}}{p} \bsym{\nabla} \widetilde{\Theta}
\end{align}
Based on these considerations, the PLT method will be applied to the considered topology optimisation problem by modifying \Cref{alg: one-shot} in the following ways:
\begin{enumerate}
    \item PLT is used instead of Parareal in Line \ref{line: one-shot parareal primal} and \ref{line: one-shot parareal adj}. 
    \item The objective and sensitivities being estimated by PLT are $\widetilde{\Theta}$ and $\bsym{\nabla} \widetilde{\Theta}$ instead of $\Theta$ and $\bsym{\nabla} \Theta$.
    \item The objective and sensitivities passed to the MMA subroutine are $\Theta$ and $\bsym{\nabla} \Theta$, and they are evaluated using \Cref{eq: obj from modded obj,eq: nabla obj from modded nabla obj}. 
\end{enumerate}
The resulting method is implemented in MATLAB and parallelised in the same way as the one-shot Parareal method. It is applied to the test case defined in \Cref{sec: test case def}. All parameters of the tests are set equal to the parameters chosen for the tests described in \Cref{sec: one-shot parareal exp procedure}, meaning that 14 tests of the PLT method are executed. The true objective is evaluated at the end of each test, and during the specific tests where $N_\tau = 5$, 10, and 30, the true objective is computed and stored at every iteration. The time taken to compute the true objective is excluded when evaluating the speedup. The reference method which the results are compared against is the same as the reference method defined in \Cref{sec: one-shot parareal exp procedure}. The tests are executed on the same type of machine as the one used to test the one-shot Parareal method.


\subsection{Results of tests of Parallel Local-in-Time method}

\begin{figure}[phtb]
    \centering
    \includegraphics[width=0.9\linewidth]{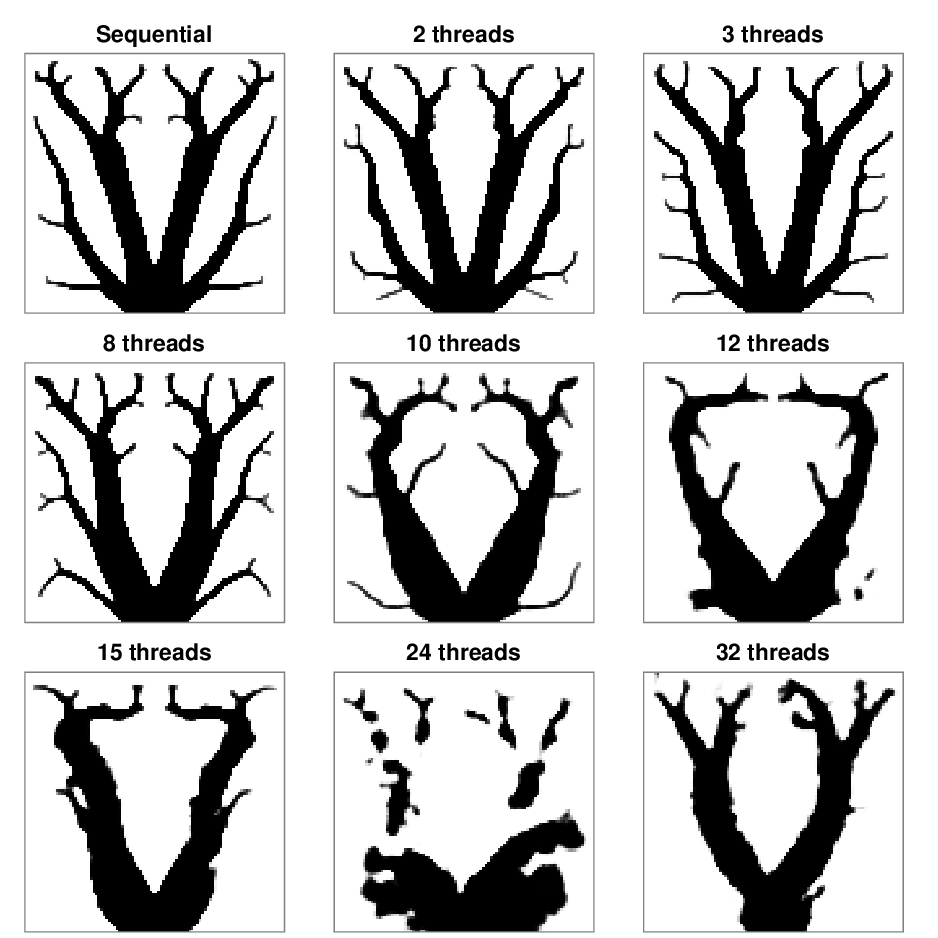}
    \caption{A selection of nine different topologies that were found using the sequential reference method and the PLT method with different numbers of threads. The PLT method is clearly unstable when using large numbers of threads. }
    \label{fig: PLT nine designs}
\end{figure}

The PLT method proved to be highly unstable for large values of $N_\tau$. This is illustrated in \Cref{fig: PLT nine designs}, which shows a sample of eight of the designs that were found at the last iteration of the tests of the PLT method, along with the reference method. The designs obtained using 8 or fewer threads look qualitatively similar to each other, but the designs begin to diverge from each other when 10 or more threads are used. This observation is consistent with the recommendation that $\Delta \tau = t_T / N_\tau$ should not be significantly smaller than the characteristic time scale of the physics when using the PLT method \cite{Theulings_PLT}. 

\begin{figure}[phtb]
    \centering
    \includegraphics[width=0.8\linewidth]{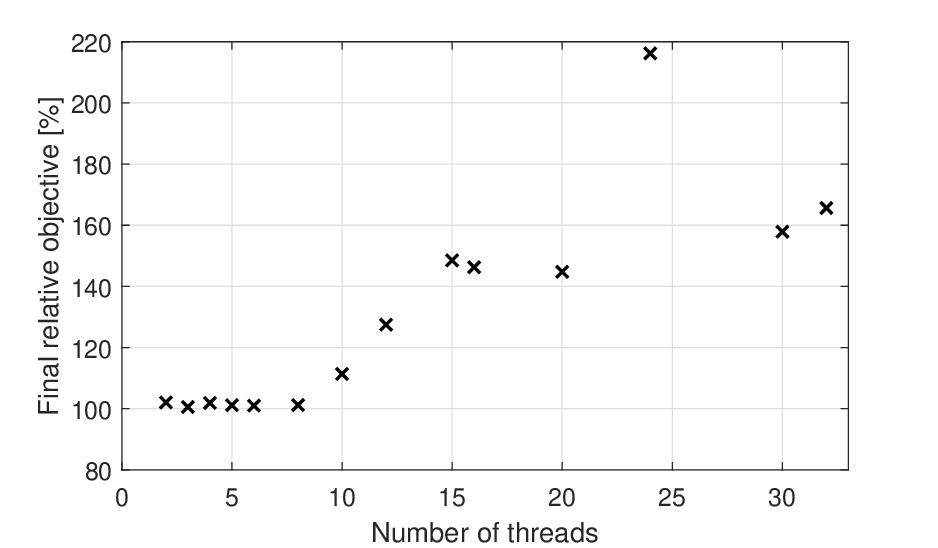}
    \caption{Relative objective values of the designs obtained at the final iterations of the tests of the PLT method of topology optimisation. These are shown as a percentage of the final objective obtained by the sequential reference method.}
    \label{fig: PLT final obj}
\end{figure}

\Cref{fig: PLT final obj} shows the relative final objective of each of the tests of the PLT method, given as a percentage of the final objective found by the reference method. It is seen that when fewer than 10 threads are used, the final objective is nearly constant and very close to the objective found by the reference method. When using 10 or more threads, the objective becomes significantly larger, and there is a positive correlation between the objective and the number of threads.

\begin{figure}[phtb]
    \centering
    \includegraphics[width=0.8\linewidth]{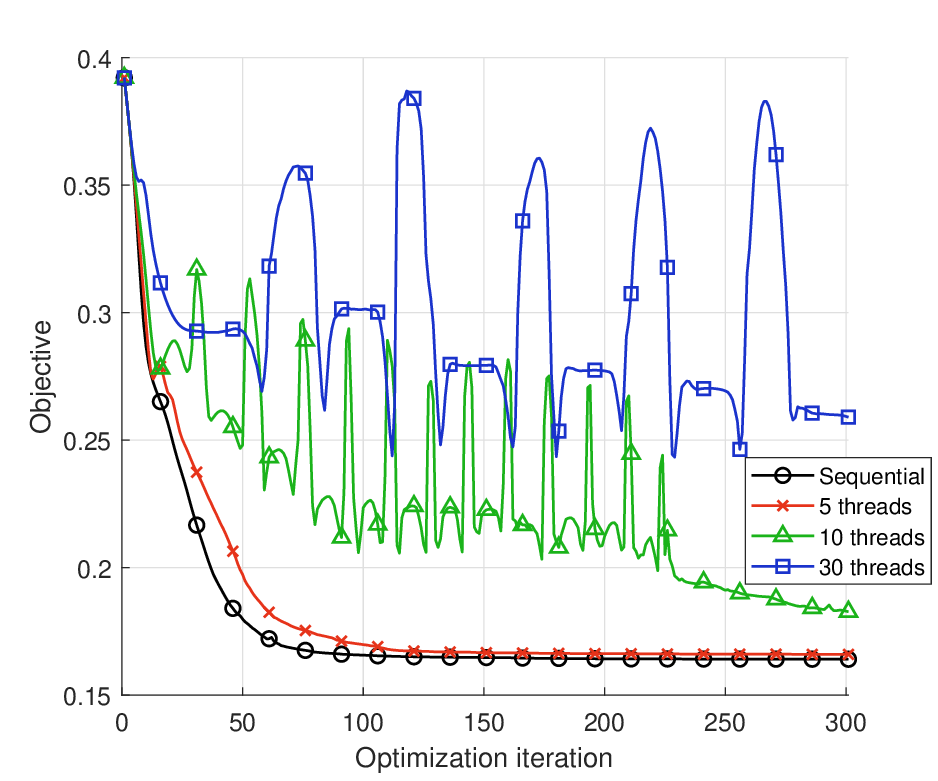}
    \caption{True objective values (as opposed to objective values estimated by PLT) as a function of the optimisation iteration during a small selection of the tests of the PLT method, along with the test of the sequential reference method. }
    \label{fig: PLT obj hist}
\end{figure}

\Cref{fig: PLT obj hist} shows the histories of the true objective for the tests where $N_\tau = 5$, 10, and 30, along with the reference method. Unlike in the tests of the one-shot Parareal method, these histories are easy to distinguish from each other. It is seen that the PLT method is fairly stable for $N_\tau = 5$, with the exception of one small spike in the objective occurring around the 15\tss{th} iteration. However, it does converge noticeably slower than the reference method, at least in terms of objective reduction per iteration. Meanwhile, in the test where $N_\tau = 10$, the objective is clearly unstable during the first 200 iterations, but it appears to stabilise around the 230\tss{th} iteration, after which it converges very slowly. In the test where $N_\tau = 30$, the objective is unstable during the entire optimisation process. 

\begin{figure}[phtb]
    \centering
    \includegraphics[width=0.8\linewidth]{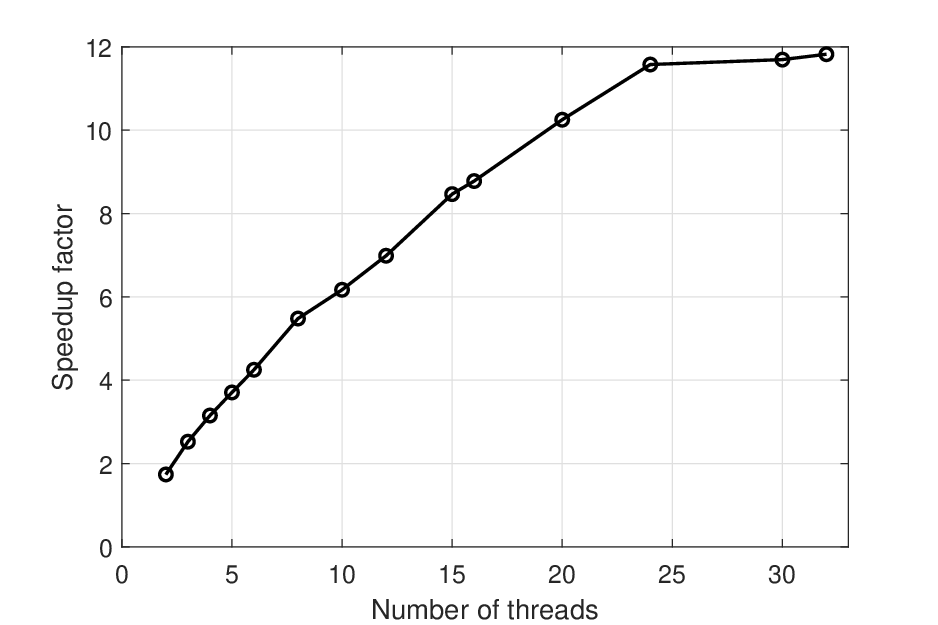}
    \caption{Measured speedup of the PLT method of topology optimisation. The peak is at 32 threads with a speedup of $ 11.8 \times $.}
    \label{fig: PLT speedup}
\end{figure}

The measured speedup of the PLT method is plotted in \Cref{fig: PLT speedup}. The peak speedup is at $11.8 \times$ using 32 threads, which is significantly better than the maximum speedup that was measured for the one-shot Parareal method, which was $4.95 \times$. But it is argued that the speedup factors measured for $N_\tau > 8$ are meaningless, since the PLT method is unstable in those cases. However, it is worth investigating why the PLT method is so much more efficient than the one-shot Parareal method, since this may lead to ideas for improvements of the one-shot Parareal method.

\subsection{Isolating bottleneck in one-shot Parareal method}


Ignoring considerations about accuracy and stability, the PLT method has the following advantages over the one-shot Parareal method in terms of computational efficiency:
\begin{enumerate}
    \item The PLT method does not send the temperature field at the intermediate time steps to the master thread. This reduces communication delays. 
    \item The parallel computations of the PLT method are not interrupted when it switches from solving the primal problem to the adjoint problem, because the primal and adjoint problems are solved in the same block of parallel code in the PLT method. This is in contrast to the one-shot Parareal method where the primal and adjoint problems are solved in separate calls to the Parareal algorithm. This further reduces communication delays for the PLT method. 
    \item The PLT method does not spend time on computing coarse propagators sequentially. 
\end{enumerate}
To investigate which of these advantages are the most significant, a test is performed where the one-shot Parareal method is modified such that $\mathcal{G}_{pri}$ and $\mathcal{G}_{adj}$ always return $\mbf{0}$. It is then executed as described in \Cref{sec: one-shot parareal exp procedure} with $N_\tau$ set to 32. 
It failed to converge within the 300 iterations given. However, the speedup relative to the reference method is measured to be $10.8 \times$. This is close to the speedup of $11.8 \times$ achieved by the PLT method for $N_\tau=32$ and significantly faster than the speedup of $3.81 \times$ measured for $N_\tau=32$ using the original version of the one-shot Parareal method. 

Based on these observations, it is evident that the most significant bottleneck for the original one-shot Parareal method is the time spent on evaluating the coarse propagators. This makes sense, since this part remains sequential and, thus, independent of parallelisation. Therefore, if any significant improvements on the speed of the one-shot Parareal method are to be made, then it would be prudent to focus on reducing the compute time of the coarse propagators. This could be done by, for example, evaluating the coarse propagators on a coarse spatial mesh or by parallelising the computation with respect to space. 
However, this is left as subject for future investigations.


\section{Conclusion}
\label{sec: conclusion}


This paper has proposed a one-shot method which uses the iterative parallel-in-time method Parareal to accelerate topology optimisation of transient heat conduction problems. In order to accommodate the adjoint sensitivity analysis, the Parareal algorithm was modified to save the temperature field at all intermediate time points. This modification came at the cost of greater memory usage and greater communication overhead. 

The method estimates the objective by introducing cumulative objectives at each time point and having Parareal solve for the cumulative objectives while solving for the temperature. Likewise, it estimates the sensitivities by introducing cumulative sensitivities and having Parareal solve for them while solving the adjoint problem. The estimates of the objective and sensitivities are then passed to the Method of Moving Asymptotes to update the design. 

Preliminary tests were made to evaluate the speed and accuracy of the above Parareal method. These revealed that Parareal could reach a speedup factor up to $5 \times$ for the considered test problem, if only one Parareal iteration was executed. However, multiple iterations were required in order to obtain accurate results, even when the imposed tolerance was very lenient. As a consequence, the speedup was lowered significantly and became useless in practise. To mitigate the large errors and low speedups, the proposed optimisation method uses warm restarts and only one Parareal iteration to solve the primal and adjoint problems during each optimisation iteration. This makes it a one-shot method. 

The one-shot Parareal method was tested by applying it to a topology optimisation problem, where the objective functional was a power mean of the temperature. The method achieved a peak speedup of $4.95 \times$ using 16 threads. The method converged to different local minima depending on the number of threads that were used. However, these local minima were all similar to the minimum found by a sequential reference method. The objective values were within $\pm 2 \%$ of the reference value and the topologies looked qualitatively similar to each other. The one-shot Parareal method did converge more slowly than the reference method in terms of objective improvement per iteration. However, this is likely offset in practise by the speedup of the one-shot Parareal method. 

It was argued that the proposed method is, in theory, similar to another parallel-in-time method of topology optimisation named the Parallel Local-in-Time (PLT) method. Tests were made to compare the performance of the PLT method to the one-shot Parareal method. This revealed that the PLT method is highly unstable when using large numbers of threads. However, it was also significantly faster than the one-shot Parareal method, reaching a speedup of up to $11.8 \times$. 

It was determined that the most significant bottleneck in the one-shot Parareal method was the time spent on evaluating the coarse propagators, due to this remaining a sequential process. This means that if significant improvements are to be made, then it makes sense to make the coarse propagators cheaper to evaluate.

\appendix


\section{Quirks regarding timing of sequential time-steppers in MATLAB}
\label{sec: timing quirks}


During some preliminary tests of the MATLAB code used in this project, it was sometimes found that it achieved a speedup greater than 2, when running Parareal using a parallel pool containing only 2 worker threads, which should not be possible. After troubleshooting, it was found that the backward Euler method ran slower when being evaluated by the master thread than when it was being evaluated by a worker thread. The performance of the sequential time-stepping code was (prior to this observation) measured by running it on the master thread, thus giving the sequential code an unfair disadvantage, which explains the false appearance of achieving an efficiency greater than 100\%. 

The authors do not know the cause of this slowdown of the master thread. However, they hypothesise that the master thread might be using a sub-optimal parallelisation of the linear solver used by the backward Euler method, because it was observed that the master thread used more CPU than an individual worker thread would use to solve the same problem. 

To obtain a more fair comparison between the performance of Parareal versus the sequential time-stepping method, it was decided to always evaluate the performance of the sequential method by creating a parallel pool consisting of only one worker thread, and then to make that one worker run the sequential method. This work-around adds a bit of uncertainty on the measured wall-clock times of the sequential code, because it is difficult to tell how much of the measured time is due to processing and how much of it is due to communication delays between the master thread and the one worker. 

It was also decided to let a worker thread run the MMA subroutine which updates the design variables, because this was also observed to be faster than letting the master thread run it.

\bibliographystyle{siamplain}
\bibliography{references}

\end{document}